\documentclass[prx,twocolumn,amsmath,amssymb,longbibliography]{revtex4}

\usepackage{graphicx}% Include figure files
\usepackage{dcolumn}% Align table columns on decimal pointb
\usepackage{bm}% bold math
\usepackage{gensymb}
\usepackage{multirow}
\usepackage[none]{hyphenat}
\usepackage{epstopdf}
\usepackage{float}
\usepackage{subfigure}
\usepackage{tabularx}
\usepackage{color}
\usepackage{siunitx}
\usepackage{xr}
\usepackage{amsmath} 
\usepackage{mathtools}
\usepackage[normalem]{ulem}
\usepackage{natbib}
\usepackage[caption=false]{subfig}
\usepackage{afterpage}
\usepackage{enumerate}
\makeatletter
\newcommand*{\rom}[1]{\expandafter\@slowromancap\romannumeral #1@}
\makeatother

\begin{document}

\newcommand{\beq}{\begin{equation}}
\newcommand{\eeq}{\end{equation}}
\newcommand{\vS}{\vec{S}}
\newcommand{\ba}{\mathbf{a}}
\newcommand{\hx}{\hat{\mathbf{x}}}
\newcommand{\hy}{\hat{\mathbf{y}}}
\newcommand{\rr}{\mathbf{r}}
\newcommand{\kk}{\mathbf{k}}
\newcommand{\qq}{\mathbf{q}}
\newcommand{\cdag}[1]{c^\dagger_{#1}}
\newcommand{\cop}[1]{c_{#1}^{\phantom{\dagger}}}
\newcommand{\an}[1]{{\color{red} #1}}
\newcommand{\re}{\mathrm{Re}}
\newcommand{\im}{\mathrm{Im}}
\newcommand{\ud}{\mathrm{d}}
\newcommand{\up}{\uparrow}
\newcommand{\dn}{\downarrow}
\newcommand{\pd}{\partial}

\title{Stability of a Nonunitary Triplet Pairing  on the Border of Magnetism in UTe$_2$}

\author{Andriy H. Nevidomskyy}
%\email[Correspondence e-mail address: ]{nevidomskyy@rice.edu}
\affiliation{Department of Physics and Astronomy, Rice University, Houston, TX 77005, USA}

\date{\today}

\begin{abstract}
Motivated by the recent discovery of superconductivity in UTe$_2$, we analyze the stability and nodal structure of various triplet superconducting order parameters. Using a combination of symmetry group-theoretical analysis, phenomenological Landau free energy and weak-coupling BCS theory, we show that chiral nonunitary superconducting order can be stabilized on the border of ferromagnetism in UTe$_2$, even in the absence of long-range magnetic order. We further perform first principles density functional theory (DFT) calculations of the so-called ``small" Fermi surface, excluding the contribution of U $f$-electrons, and find it to be in excellent agreement with the recent angular resolved photoemission study and DFT+DMFT calculations. This permits us to elucidate the nodal structure of the superconducting gap, which we find generically to possess point nodes along the crystallographic $a$ direction, in agreement with experiments. The topological stability of these point nodes and their associated Majorana surface states is analyzed. The nonunitary structure of the predicted superconducting state supports chiral edge modes observed in recent scanning tunneling microscopy (STM) data and is predicted to result in a non-vanishing magneto-optical Kerr effect. 

\end{abstract}

\maketitle

%\section{Introduction}

Recent discovery of superconductivity in UTe$_2$~\cite{Ran19,Aoki19} has attracted much attention, inviting comparisons with ferromagnetic superconductors based on other uranium compounds UGe$_2$, UCoGe, URhGe and UCoAl~\cite{Aoki14,Aoki19-review}. Indeed, a large upper critical field has been found to exceed the Pauli--Clogston limit, as expected from the spin-triplet superconductivity, also corroborated by the very weak temperature dependence of the nuclear magnetic resonance (NMR) Knight shift below $T_c$~\cite{Ran19,Nakamine-NMR19}. In contrast to ferromagnetic (FM) U-based superconductors however, no long-range magnetic order has been conclusively identified, although recent pressure measurements indicate~\cite{Braithwaite19,Ran-pressure} a possible onset of a FM phase above a critical pressure $P_m \approx 1.4 \div 1.7$~GPa. It is therefore tempting to conclude that superconductivity in UTe$_2$ is likely mediated by critical magnetic fluctuations on the brink of the ferromagnetic transition. 

At the same time, several outstanding issues remain in our understanding of this material. Because the orthorhombic $D_{2h}$ point group symmetry only allows one-dimensional representations, it has been suggested that the superconducting (SC) order parameter must be unitary, not breaking the time-reversal symmetry (TRS). On the other hand, the appearance of re-entrant superconductivity in large applied fields~\cite{Ran19-fields,Knebel19} and the recent scanning tunneling microscopy (STM)~\cite{Madhavan19} provide evidence for TRS-breaking chiral superconductivity, which would be consistent with nonunitary order parameter~\cite{Ran19}. In light of the above, the nodal structure of the SC order parameter is also not fully determined, with several unitary orders that are predicted to be fully gapped, while the power-law temperature dependence of the thermal conductivity, specific heat and penetration depth point towards the point nodes in the gap~\cite{Metz19}. Complicating the analysis is the apparent discrepancy between the metallic nature of the material in the experiment and the insulating behavior seen in the \textit{ab initio} density functional theory (DFT) calculations in the paramagnetic phase~\cite{Aoki19,Shick19}, prompting some authors to paint this as an insulator-metal transition with the Coulomb interaction inducing, counter-intuitively, a metallic state~\cite{Ishizuka19}. The geometry of the Fermi surface is debated~\cite{Fujimori19,Xu19,Ishizuka19}, leading to uncertainty in establishing the nodal structure of the SC gap.

In this paper, we analyze the possible pairing structure of UTe$_2$ and its reported coexistence with ferromagnetism under the applied pressure~\cite{Ran-pressure}. Using a combination of phenomenological free-energy analysis and the weak-coupling Bardeen--Cooper--Schrieffer (BCS) theory, we propose that a nonunitary triplet superconducting order parameter becomes energetically most favourable even in the paramagnetic phase, due to the reported proximity to ferromagnetic order. This nonunitary state, which spontaneously breaks the time-reversal symmetry, is also stabilized by the external magnetic field ($H>H_{c1}$), which qualitatively agrees with the observation of re-entrant superconductivity in high fields~\cite{Ran19-fields,Knebel19}. Our analysis further explains the stability of the nonunitary order parameter even in the absence of long-range order or inherent magnetization. The chiral nature of the proposed nonunitary state supports the recently found evidence for chiral edge currents in the STM study~\cite{Madhavan19}. We further predict that because of the spontaneously broken TRS in the nonunitary state, it ought to have a non-vanishing magneto-optical Kerr response. 

In order to elucidate the nodal structure of the gap, we address the aforementioned dichotomy among the different DFT and DFT+$U$ calculations on UTe$_2$.  We compute the ``small" Fermi surface without the $f$-electron contribution and find that it is in excellent agreement with the DFT+DMFT calculations~\cite{Xu19} and with recent angular resolved photoemission spectroscopy (ARPES) study~\cite{Miao-ARPES19}. Inspiring further confidence in this treatment is the fact that large applied magnetic fields~\cite{Ran19-fields,Knebel19} and/or the inherent magnetization due to pressure-induced FM order~\cite{Braithwaite19,Ran-pressure} are expected to quench the Kondo screening of U localized moments by conduction electrons, thus  justifying the analysis of the SC gap on the ``small" Fermi surface due to conduction electrons alone.  
%We extend this line of reasoning to the strong magnetic fields ($H_{c1}\! <\! H\! <\! H_{c2}$), which may also drive the non-unitary pairing in UTe$_2$.

%\section{
\textit{Stability of a nonunitary SC order.}
The triplet superconducting gap in the limit of strong spin-orbit coupling is characterized by the $\mathbf{d}$-vector $\Delta(\kk) =   i(\vec{d}_\kk\cdot \vec{\sigma}) \sigma_2$, which transforms under the irreducible representations (irreps) $\Gamma_i(\hat{\kk})$ of the point-group symmetry
$\vec{d}_\kk = \sum_{i} d_{\Gamma_i} \Gamma_i(\hat{\kk}).$
In the case of UTe$_2$, the point group $D_{2h}$ contains only the one-dimensional representations, which would generically have different transition temperature $T_c$~\cite{Yip93}. Therefore, it was argued~\cite{Xu19} that the SC order parameter in UTe$_2$ must be single-component. This conclusion however seemingly fails to explain the recent experimental evidence of chiral edge currents seen by the %scanning tunneling microscopy 
(STM)~\cite{Madhavan19} or the residual value of specific heat below $T_c$ at half the normal state value~\cite{Ran19}. 

%Let us consider an internal magnetization $\textbf{M}$ induced in this type-II superconductor either by an applied magnetic field above $H_{c1}$, or imposed by an intrinsic ferromagnetic order, which may develop under the applied pressure~\cite{Braithwaite19,Ran-pressure}.  
It has been argued that applying the magnetic field in, say, crystallographic $\hat{b} \parallel \hat{y}$ direction (where $H_{c2}$ is largest~\cite{Ran19}), changes the point group symmetry to $C_{2h}^y$, allowing for a linear superposition of the two odd irreducible representations $B_{1u}$ and $B_{3u}$~\cite{Ishizuka19}. The lowest-order in $|k|$ characteristic functions of these two irreps are given by
\begin{align}
    B_{1u}: \vec{d}_1(\kk) & = \eta_b k_a \hat{b} + \eta_a k_b \hat{a}   \nonumber\\
    B_{3u}: \vec{d}_3(\kk) & = \eta_b' k_c \hat{b} + \eta_c k_b \hat{c}, \label{eq:irreps}
\end{align}
where the constants $\eta_i$ are assumed to be real. The other two irreps, $A_u$ and $B_{2u}$, are deemed less favorable because they are either fully gapped ($A_u$), or expected to have point nodes along the $b$-axis ($B_{2u}$), contrary to thermal conductivity measurements~\cite{Metz19}. We therefore  focus on $B_{1u}$ and $B_{3u}$ states in what follows.
Let us consider a superposition of these two order parameters written, without loss of generality, as $\alpha B_{1u} + iB_{3u}$. If $\alpha$ is taken to be purely imaginary, this would be the unitary $B_{1u} + B_{3u}$ state considered in Ref.~\onlinecite{Ishizuka19}. If, on the other hand, $\re(\alpha)\neq 0$, %has a real component,
that would result in a chiral \textit{nonunitary} state. %with a non-vanishing orbital component, as will be shown below. 
The $\vec{d}$-vector of such a state takes the form
\beq
\vec{d}(\kk) \sim \alpha\eta_a k_b \hat{a} + (\eta_b \alpha k_a + i \eta_b' k_c) \hat{b} + i\eta_c k_b \hat{c}.
\label{eq:dvec}
\eeq
%where we have taken the coefficient $\eta_y' = \eta_y$ in Eqs.~(\ref{eq:irreps}), without loss of generality, to lighten the notation.

The orbital moment of a Cooper pair is defined as~\cite{Machida01}
\begin{align}
\mathbf{m}_\text{orb}(\kk) & = i\vec{d}_\kk \times \vec{d}^*_\kk \label{eq:morb}\\
             & = 2\re(\alpha) \left[ \eta_{bc}' k_a k_b\, \hat{a} -  \eta_{ac} k_b^2\, \hat{b} + \eta_{ab} k_b k_c\, \hat{c} \right], \nonumber 
\end{align}
where we used the shorthand $\eta_{\alpha\beta} \equiv \eta_\alpha \eta_\beta$.
Application of an external field ($H>H_{c1}$ in a type-II superconductor) would result in the internal magnetization $\mathbf{M}=\chi \mathbf{B}$ ($\chi$ is the normal state susceptibility), which will then couple to  
%$\mathbf{B} = \mu_0 \mathbf{H} + \mathbf{M}$, 
the orbital moment of the Cooper pair, resulting in the contribution to the system free energy:
\beq
\Delta F_\text{orb} =w\iint\limits_{FS} \ud^2 k\;  \mathbf{m}_\text{orb}(\kk) \cdot \mathbf{M},
\label{eq:forb}
\eeq
where the integral is over the Fermi surface and $w$ is a phenomenological coupling constant.
%where the coefficient $w \propto |\Delta_0|^2$ is proportional to the overall magnitude of the superconducting gap. 
The Landau free energy describing the coexistence of superconductivity and fluctuating magnetization is hence of the form (we drop integration over $\kk$ for brevity)
\begin{align}
    F & =  a|\vec{d}|^2 + b |\vec{d}|^4 + c |\vec{d}\times \vec{d}^\ast|^2 + w \mathbf{M}\cdot(i\vec{d}\times  \vec{d}^\ast) \nonumber \\
    & +  \frac{|\mathbf{M}|^2}{2\chi(T)} + u|\mathbf{M}|^4 - \mathbf{H}\cdot \mathbf{M}.
    \label{eq:free}
\end{align}
We see that the orbital momentum of the Cooper pair leads to the additional contribution to magnetization $\Delta \mathbf{M} = -\chi w (i\vec{d}\times  \vec{d}^\ast)$. 
%$\Delta \mathbf{M} = -\chi w (i\vec{d}\times  \vec{d}^\ast) \propto T_c - T$, since $|\vec{d}| \propto \sqrt{T_c - T}$ below the superconducting $T_c$. 
Substituting into Eq.~(\ref{eq:free}) yields
\beq
F_{\text{eff}}[\vec{d}(\kk)] = a|\vec{d}|^2 + b |\vec{d}|^4 + \left(c -\frac{\chi w^2}{2}\right) |\vec{d}\times \vec{d}^\ast|^2,
\label{eq:free2}
\eeq
and we see that even if nonunitary state is initially not favoured ($c > 0$), the coupling to magnetization can stabilize it if $(c-\chi w^2/2) <0$, especially as the normal state susceptibility $\chi \sim (P-P_m)^{-1}$ diverges upon approaching the magnetic critical point as a function of, say, pressure $P$. The value of the coefficient $\alpha$ entering the $\vec{d}$-vector is then chosen variationally such as to minimize the free energy Eq.~(\ref{eq:free2}). This underlines our main conclusion -- the effect of coupling to fluctuating magnetization, \textit{even inside the PM phase}, is to induce a finite value of $\re(\alpha)$, resulting in a chiral nonunitary SC state of the form $B_{1u} + iB_{3u}$~\cite{endnote1}.

\textit{Weak-coupling BCS theory.}
The upper critical field $H_{c2}$ is very anisotropic in UTe$_2$, with the largest value for $H\parallel \hat{b}$~\cite{Ran19}. Motivated by this and recent high-field experiments~\cite{Ran19-fields}, we shall therefore assume that the field-induced magnetization $\mathbf{M} \parallel \hat{b}$ (for general case of arbitrary field direction, see Supplementary Materials~\cite{SM}).
It is thus convenient to choose a coordinate systems such that $\hat{z} \parallel \hat{b}$ is along the internal magnetization. This is achieved by the SO(3) rotation of the coordinates such that $(a, b, c) \to (y, z, x)$. In this coordinate system, the $\vec{d}$-vector in Eq.~(\ref{eq:dvec}) results in the order parameter $\hat{\Delta}=(i\vec{\sigma}\sigma_2\cdot \vec{d}_\kk)$  of the form
\beq
\hat{\Delta}(\kk) = i\left( \begin{array}{ccc}
 -\eta_c k_b + \alpha \eta_a k_b   & & \eta_b' k_c - i\alpha \eta_b k_a  \\
 \eta_b' k_c - i\alpha \eta_b k_a   & &  \eta_c k_b + \alpha \eta_a k_b
\end{array}\right).
\label{eq:Delta-full}
\eeq
Under application of a strong magnetic field, or in a FM state possibly induced by pressure~\cite{Braithwaite19}, the strong Zeeman splitting of the Fermi surfaces will result in the Fermi-vector mismatch between the majority and minority Fermi surface sheet. This will therefore suppress the off-diagonal matrix elements of the pairing amplitude in Eq.~(\ref{eq:Delta-full}) -- the situation described as \textit{equal-spin pairing}~\cite{endnote2}.
%~\endnote{Making the assumption of equal-spin pairing, which is to say suppressing the $z$-component of the $\vec{d}$-vector, also has a clear physical meaning even in the PM phase, where the Zeeman splitting due to applied field may be small -- setting $d_z=0$ ensures that the transverse components of the orbital moment vanish, $m_x \propto \im(d_z d_y^\ast) = 0$ and $m_y \propto \im(d_z d_x^\ast) = 0$, so that the component of the orbital moment along the field direction, $m_z$, is maximized.}.  
We can see from Eq.~(\ref{eq:Delta-full}) that in order for this to occur, we should set $\eta_b' = 0 = \eta_b$. The resulting matrix $\Delta_{\alpha\beta} = \mathrm{diag}\{-\Delta_-, \Delta_+\}$ describes the pairing on the two Zeeman-split Fermi surfaces, with $\Delta_{\pm}(\kk) = (\eta_c \pm \alpha \eta_a) k_b$. 

The values of the superconducting pairing amplitudes on the two spin-split Fermi surfaces are determined from the coupled BCS equations (written at zero temperature, for simplicity):
\begin{align}
    \Delta_- & = -\sum\limits_{\kk'} V_{\up\up}(\kk,\kk') \frac{\Delta_-(\kk')}{2E_-}  - V_{\up\dn}(\kk,\kk') \frac{\Delta_+(\kk')}{2E_+} \label{eq:BCS1}  \\
     \Delta_+ & = -\sum\limits_{\kk'} V_{\dn\dn}(\kk,\kk') \frac{\Delta_+(\kk')}{2E_+}  - V_{\dn\up}(\kk,\kk') \frac{\Delta_-(\kk')}{2E_-},\label{eq:BCS2}
\end{align}
with  the Bogoliubov spectrum given by $E_{\pm}(\kk') = \sqrt{\xi_\pm(\kk')^2 + |\Delta_{\pm}(\kk')|^2}$ in terms of the Zeeman split normal state energies $\xi_\pm(\kk)=\xi_0(\kk) \pm g\mu_B B$ (field \mbox{$B= \mu_0H + M$} includes internal FM moment, if present). 

In order to make progress in solving Eqs.~(\ref{eq:BCS1}--\ref{eq:BCS2}), one would need a microscopic model for the pairing, which is  difficult to establish based on current experimental evidence. One may entertain several options.

\textit{Electron-phonon pairing.}
Normally, electron-phonon pairing would favor $s$-wave superconductivity, however in the presence of external magnetic field above the Pauli--Clogston limit, or when the SC order coexists with ferromagnetism, electron-phonon mechanism can result in odd-wave pairing. In the first approximation, coupling to phonons preserves the pseudo-spin of electrons in the Cooper pair~\cite{endnote3},
%\endnote{The electron spin is not conserved because of the spin-orbit coupling, however the pseudospin, defined here as the projection of total angular momentum, is a good quantum number.}, 
so that we may approximate $VB$, reducing Eqs.~(\ref{eq:BCS1}--\ref{eq:BCS2}) to two  independent gap equations. Further, the pairing potential is independent of the spin, $|V_{\sigma\sigma}| = V$, and the sole reason one expects to obtain different values for $\Delta_+$ and $\Delta_-$ is because of the Zeeman splitting of the Fermi surfaces, inducing a different density of states $N_+(0) \neq N_-(0)$, so that $\Delta_{\pm} = 2\omega_D \exp[-1/(V N_\pm(0))]$ has two solutions ($\omega_D$ is the phonon Debye frequency). 

\textit{Pairing due to magnetic fluctuations.}
The exchange of damped spin waves (so-called paramagnons) in the vicinity of a FM phase transition has been long considered as a source of SC pairing in $^3$He~\cite{Berk-Schrieffer,Anderson-Brinkman} and in ferromagnetic superconductors alike~\cite{Fay-Appel,AN2005}. The magnitude of the pairing potential, computed within the random phase approximation (RPA), is given by~\cite{Fay-Appel}
\beq
V_{\sigma\sigma} = -\frac{I^2 \chi_{\bar{\sigma}}}{1-I^2\chi_\sigma \chi_{\bar{\sigma}}},
\eeq
where $\chi_{\sigma}$ is the spin susceptibility (Lindhardt function) for a given spin $\sigma$ ($\bar{\sigma}\equiv -\sigma$), and $I$ is the strength of the contact (Stoner) interaction. The expression for $V_{\up\dn}$ is more complicated and is given in \cite{SM}. In zero field in PM phase, it can be shown that for odd pairing $V_{\up\dn} = V_{\up\up}$, so that its contribution to Eqs.~(\ref{eq:BCS1}--\ref{eq:BCS2}) cannot be neglected~\cite{Fay-Appel}. Moreover, $V_{\up\dn}$ is strongly pair-breaking in the even parity (spin-singlet) sector, thereby favouring odd-wave pairing. In the FM phase, $V_{\up\dn}\neq V_{\sigma\sigma}$ generically, and the details depend on the Fermi surface geometry and the strength of the Zeeman splitting due to the FM order.

\textit{Consequences for specific heat.} Above we showed that even in the PM phase, the nonunitary SC state can be stabilized (see Eq.~\ref{eq:free2}) with nonzero orbital moment of the Cooper pair. The latter can be expressed as 
\beq
m_z(\kk) = \frac{1}{2}\left(|\Delta_-|^2 - |\Delta_+|^2\right) = -2\alpha\eta_a\eta_c k_b^2,
\eeq
meaning that the pairing amplitude is different for the two spin components. Let us take $|\Delta_-| < |\Delta_+|$, without loss of generality, which would imply two transition temperatures $T_c^{-} < T_c^{+}$, with the upper identified as the bulk $T_c$.
Thus, \textit{regardless of the pairing mechanism}, there will be a range of temperatures below $T_c$ where the minority Fermi surface will remain gapless, contributing to the specific heat in the normal state, precisely as observed in UTe$_2$, where half of the normal specific heat remains below $T_c$~\cite{Ran19}. Inside a FM phase, this mechanism is well known and was used to explain SC in, for instance, UGe$_2$~\cite{Machida01,Linder08}. The novelty here is the applicability of the same phenomenology  in the PM phase. 

%Note that as long as the Fermi surfaces are Zeeman split, one expects $|\Delta_+| \neq |\Delta_-|$, regardless of the detailed pairing mechanism. As a consequence, Cooper pairs will carry the orbital momentum along the field direction ($\hat{b}$ axis): $m_z = (|\Delta_-|^2 - |\Delta_+|^2)/2 = -2\alpha\eta_a\eta_c k_b^2$, in agreement with Eq.~(\ref{eq:morb}). 

%\section{Nodal structure of the pairing}
\textit{First principles calculations.}
In order to elucidate the nodal structure of the superconducting pairing, we need information on the geometry of the Fermi surface. In the absence of quantum oscillation measurements, this information has been extracted from first principles calculations in earlier works on UTe$_2$~\cite{Aoki19,Shick19,Fujimori19,Xu19,Ishizuka19}. What transpired is that within the conventional local density approximation (LDA) or the generalized gradient approximation (GGA), paramagnetic UTe$_2$ turns out to be an insulator~\cite{Aoki19,Shick19} with the spectral gap induced by the $f$-$c$ hybridization between the uranium $f$-orbitals and conduction electrons, or a metal with a very small Fermi surface~\cite{Fujimori19}. While careful treatment of this problem requires solving a periodic Anderson model, which is highly non-trivial (for a dynamical mean-field theory treatment, see Ref.~\onlinecite{Xu19}), other authors have chosen to perform approximate LDA+$U$ calculations~\cite{Anisimov93}, and found that upon choosing a suitably high value of Hubbard $U$, the metallic Fermi surface appears~\cite{Xu19,Ishizuka19}.

There are obvious difficulties with this approach: the Fermi surface geometry  depends sensitively on $U$ (even as $U$ is varied slightly from 1.0 to 1.1~eV~\cite{Ishizuka19}), and more fundamentally, the LDA+$U$ method was designed to work with long-range spin order (typically antiferromagnetic), whereas  performing LDA+$U$ calculations in the PM phase simply results in a shift of the localized bands relative to the Fermi level, rather than opening up of the many-body gap between the lower and upper Hubbard bands~\cite{Anisimov93}. This does not mean that for a certain value of $U$, LDA+$U$ cannot mimic the physical Fermi surface, but such a match would be fortuitous, at best, and would lack predictive power. 

\begin{figure}[tb]
\includegraphics[width=0.48\textwidth]{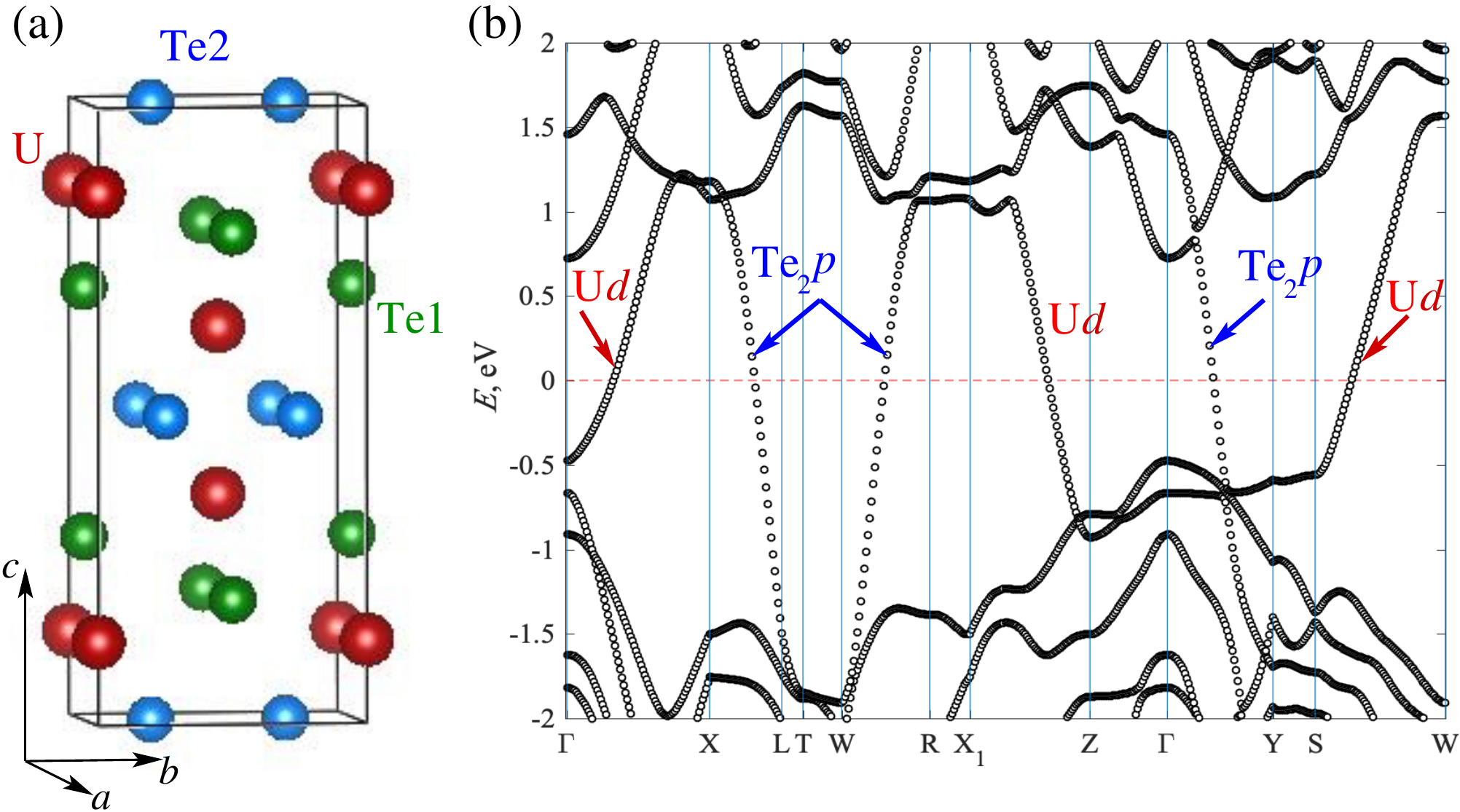}
\caption{(a) Unit cell of UTe$_2$; (b) DFT band structure plotted along the high-symmetry lines in the Brillouin zone, without $f$-electron contribution (in-core). The dominant contribution of Te $p$- and U $d$-orbitals at the Fermi level is indicated. Note that the Fermi level position has been adjusted to correspond to that in the all-electron calculation.}
\label{fig:bands}
\end{figure}

It is apparent that a different approach is needed and in this work, we instead opt for a fully \textit{ab initio} (with no adjustable parameters) treatment, by ``freezing" the $f$-orbitals of uranium into the core. Such a frozen-core calculation is common for analyzing $f$-electron systems and while it obviously ignores the important role played by the localized orbitals, it does have an advantage of elucidating what is often referred to as ``small" Fermi surface, when the $f$-$c$ hybridization is absent and $f$-electrons do not contribute to the Fermi surface. There is a good reason to do so in UTe$_2$ when investigating the effect of strong magnetic fields or possible ferromagnetism -- in that case, the Zeeman splitting of the Fermi surface is expected to overcome the Anderson $f$-$c$ hybridization, thus denying the Kondo screening of the local moments and resulting in the ``small" Fermi surface. That this is relevant to UTe$_2$ becomes apparent from the estimates of the Kondo temperature $T_K \sim 30$~K from fitting the Fano lineshape of the STM tunneling conductance data~\cite{Madhavan19}, indicating that internal fields of the order of 30~T would be sufficient to quench the Kondo effect and reveal the underlying ``small" Fermi surface originating from conduction electrons alone. The suppression of the Kondo coherence by applied pressure of the order of $1$~GPa has also been clearly observed~\cite{Ran-pressure}.

\begin{figure}
    \centering
    \includegraphics[width=0.48\textwidth]{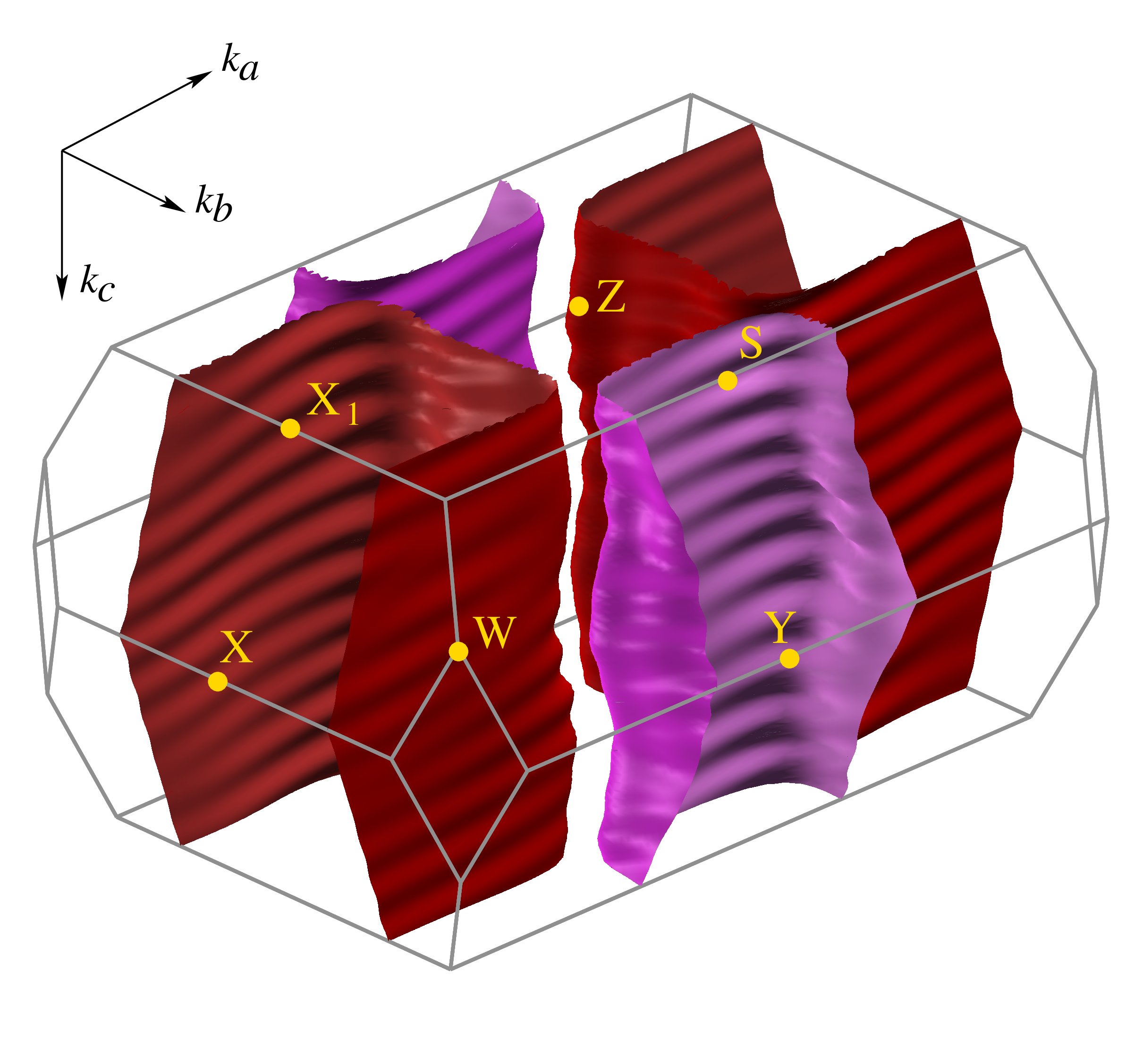}
    \caption{``Small" Fermi surface of UTe$_2$, computed without contribution of $f$-electrons. The red sheet corresponds to the hole pocket around $X$-point, and the purple sheet to the electron pocket around the $Y$ point (see bands in Fig.~\ref{fig:bands}).}
    \label{fig:FS}
\end{figure}

Figure~\ref{fig:bands} shows the band structure of the frozen-core DFT calculation~\cite{SM}. It turns out that the dominant contribution near the Fermi surface comes from conduction electrons of Te2 $p_z$-orbital and U $d$-orbitals, indicated next to the relevant bands in Fig.~\ref{fig:bands}b). These bands result in two Fermi surfaces, shown in Fig.~\ref{fig:FS}: the hole-like sheet around the $X$ point ($X$-sheet, red) and electron-like sheet around $Y$ point ($Y$-sheet, purple). The sheets are quasi-two dimensional, with essentially no dispersion along $k_c$. We note that the ``small" Fermi surface in our calculations is in excellent agreement with the recent DFT+DMFT calculations~\cite{Xu19}, inspiring further confidence in our analysis.

\textit{Nodal structure of the SC gap.}
The non-unitary equal-spin pairing described above has order parameters \mbox{$\Delta_{\pm} \propto \sin(k_b b)$} on both spin-polarized Fermi surfaces %\cite{endnote4},
\endnote{One should recall that the k-dependence quoted in Eq.~(\ref{eq:irreps}) is the lowest-order expansion in $k$ of a lattice-periodic functions, and hence the linearized dependence $\Delta\!\propto\! k_b$ should be interpreted as \mbox{$\Delta \propto \sin(k_b b)$}, leading to a pair of nodes at $k_b=0$ and $k_b=\pm \pi/b$.} 
and is expected to result in chiral surface currents, as observed recently in a recent STM experiment~\cite{Madhavan19}. Because of its $\kk$-dependence, it is expected to have zeros along the plane $k_b = 0$, which would intersect the $X$ Fermi surface sheet at a pair of line nodes $k_a = \pm k_{FX}$ parallel to $k_c$-axis, and along the place $k_b=\pi$, intersecting the $Y$-sheet at $k_a = \pm k_{FY}$. 
%The presence of such nodes along the $a$-axis has been inferred from the thermal transport~\cite{Metz19} and is consistent with the STM measurements~\cite{Madhavan19}. 
Note however that the presence of any non-zero $\eta_b'$ term in the off-diagonal elements of Eq.~(\ref{eq:Delta-full}), expected in the PM phase in the absence of Zeeman splitting, would reduce the zeros of the gap to two pairs of point nodes (rather than nodal lines) along the $a$-axis. This would agree with the expectation of point nodes from the power-law dependence of thermal conductivity and penetration depth in the PM phase~\cite{Metz19}. 

\textit{Consequences for topology.} The predicted point nodes  along the $k_a$-axis come in pairs on opposite side of the cylindrical Fermi surface, separated by $2k_F^a$. These gapless points are topologically stable in the TRS-breaking nonunitary phase, in a sense that one can define a non-zero net flux of the Berry curvature through any plane that intersects the interval between the two points and has a normal component along $k_a$:
%, for instance through $(bc)$-plane:
\beq
\Omega_a(\kk) =  \hat{n}\cdot (\pd_{k_b} \hat{n} \times \pd_{k_c} \hat{n}),
\eeq
defined in terms of the unit vector $\hat{n} = \vec{n}/|\vec{n}|$ along $\vec{n} = (\re(\Delta_\kk), \im(\Delta_\kk), \xi(\kk))$~\cite{Goswami15}. 
The pair of nodal points can thus be thought as the sources and sinks of the Berry flux, and it is well established that they  result in the gapless Majorana surface states (Majorana arcs)  joining the projections of these points onto the surface BZ~\cite{Sau-Tewari12}. A practical consequence is that one expects~\cite{Goswami15} to observe anomalous thermal Hall effect in the plane crossing these arcs, so that in our example, $\lim\limits_{T\to 0}\kappa_{bc}/T \propto C k_F^a/\pi$, where $C = \frac{1}{2\pi}\int \Omega_a(\kk) d^2 k$ is the Chern number (monopole strength) associated with these nodal points~\cite{Goswami15}.

% The integral of the Berry flux through the cross-section of the BZ defines the monopole charge (the Chern number), identical to the Shankar invariant in the $A$ phase of $^3$He~\cite{Shankar77} and to the monopole number in, e.g. the non-unitary phase of UPt$_3$~\cite{Goswami15}. It is well known that  

It is instructive to compare the non-unitary $B_{1u}+iB_{3u}$  state discussed here with the limiting cases of unitary superconductor: $B_{3u}$ has two pairs of point nodes at $k_a = \pm k_{FX}$ on the $X$-sheet of the Fermi surface, where it is crossed by two lines: $k_b=k_c=0$ line and $k_b=0, k_c=\pi$ line. 
By contrast, $B_{1u}$ state is characterized by the gap $|\Delta_{1u}|=\sqrt{\eta_a^2 k_b^2 + \eta_b^2 k_a^2}$, which only vanishes along the $\Gamma-Z$ line in the center of the BZ, which does not cross any of the Fermi surfaces. The pure $B_{1u}$ state therefore is unlikely to be realized in UTe$_2$, as it contradicts the nodal structure of the gap inferred from the experiment~\cite{Metz19}. Even if a closed pocket of the Fermi surface appears along the $\Gamma-Z$ line due to hybridization with U $f$-orbitals, it would have nodes along the $c$ direction, contrary to the thermal conductivity data~\cite{Metz19}.

%In the absence of superposition ($\alpha=0$), the two SC states characterized by the irreps $B_{1u}$ and $B_{3u}$ have line nodes in the Brillouin zone (although not necessarily on the Fermi surface), as easily seen from Eq.~(\ref{eq:irreps}). One must remember however that the structure of these expressions is the lowest-order expansion in the powers of angular momentum components $k_\alpha$. For instance, $k_x$ ought to be interpreted as $\sin(a k_x)$, which vanishes at $k_x=0$ or $\pi$ $\mod 2\pi$. As a result, the $B_{1u}$ order parameters has four line nodes given by $\kk = (0,0,k_z), (\pi,0,k_z), (0,\pi,k_z)$ and $(\pi,\pi,k_z)$, for arbitrary $k_z$. Such lines nodes generically do not cross the Fermi surface of UTe$_2$, as computed in Refs.~\onlinecite{Ishizuka2019,others}, so we conclude that the state $B_{1u}$ is fully gapped.  {\color{red} \textbf{Youichi:} is it correct that the $\Lambda$ line does not cross any FS? It's difficult to see from the picture in your paper, but Ishizuka-san can certainly tell us for sure.}

%By contrast, the $B_{3u}$ pairing vanishes along the lines $k_y, k_z = 0$ ($\Sigma$ line) or $k_y, k_z = \pi$ ($F$ line), which cross the \textit{ab initio} Fermi surfaces at pairs of points with finite $k_x = \pm K_0$. The state $B_{3u}$ therefore is characterized by point nodes on the Fermi surface. 

\textit{Unitary $(B_{1u} + B_{3u})$ state.}
While the non-unitary state discussed above is realized for $Re(\alpha)\neq 0$ in Eq.~(\ref{eq:dvec}), a unitary counterpart corresponding to purely imaginary $\alpha$ has been discussed previously in the literature~\cite{Ishizuka19}, that is the $B_{1u} + B_{3u}$ state. Its nodal structure  is governed by two conditions: (i) $k_b=0$ or $\pi$ and (ii) $k_c = -(\eta_b/\eta_b') k_a$ (we have absorbed the coefficient $|\alpha|$ into the definition of $\eta_b$). Depending on the ratio $\eta_b/\eta_b'$, the line defined by these equations may cross the $X$-sheet of the Fermi surface at a pair of nodal points (red surface in Fig.~\ref{fig:FS}), or it may avoid it entirely, leading to a fully gapped state similar to pure $B_{1u}$. Compared to the non-unitary state discussed here, such a unitary state will have a vanishing orbital momentum of the Cooper pair, and would not minimize the free energy in Eq.~\ref{eq:forb} in the presence of FM order parameter or applied field.

\textit{Field along $a$-direction}. Above, we have considered the non-unitary state stabilized by the  magnetic field along the crystallographic $\hat{b}$ direction, motivated by experiments in the applied field~\cite{Ran19-fields,Knebel19} and the fact that $H_{c2}$ is largest in the $\hat{b}$ direction~\cite{Ran19}. Analogous calculation can be performed for a field along the easy $\hat{a}$ axis, where FM order parameter could likely develop under the applied pressure~\cite{Braithwaite19,Ran-pressure}. In this case~(see Supplementary Materials~\cite{SM}), the order parameter on the two Zeeman-split Fermi surfaces is of the form
\beq
\Delta_{\pm} = (\alpha \eta_b k_a + \mp \eta_c k_b) + i\eta_b' k_c ,
\eeq
which is expected to have nodal points in the $k_c=0$ or $k_c=\pi$ plane provided the line $k_b = \pm s k_a$ crosses the Fermi surface, with the tangent given by $s=\alpha \eta_b/\eta_c$~\cite{SM}. Generically, one expects the nodes on the $X$-sheet for $|s|\lesssim 1$ and on the $Y$-sheet for $|s|\gtrsim 1$. In principle, there exists a very narrow range of $|s|\approx 1$ where the order parameter has a full gap~\cite{SM}, however this would require $\alpha$ to be fine-tuned, corresponding to a narrow range of internal field for it to be realized.

%The non-unitary state is more interesting -- it is generically fully gapped, because the Fermi surfaces, as computed in Refs.~\onlinecite{Ishizuka19,others}, do not contain the $\Gamma$ or the other 7 TRI points. It should be noted that such a state, breaking time-reversal symmetry in the presence of strong spin-orbit coupling, belongs to the Cartan class $D$, which is topologically trivial in three spatial dimensions. 
%{\color{red}\textbf{Youichi:} what can we conclude from the topological indices that you computed at the TRI points in your paper? If the SC gap is open at these 8 points, what does it tell us about the topology of such a state? Class D should be topologically trivial, in the absence of crystalline symmetries -- does the presence of $C_{2h}$ symmetry change this conclusion?}

%\vspace{10mm}
%{\color{red}\textbf{Can we make a connection to Vidya Madhavan's experiment, qualitatively? In particular, can we argue in favour of the chiral edge modes in a 2D geometry?} \\ Nevertheless, such a non-unitary state can have interesting characteristics when confined to thin films \ldots }

%\section{Conclusions}
To summarize, our calculations based on the phenomenological treatment of the Landau free energy and weak-coupling BCS analysis, reveal that the nonunitary superconducting order parameter, of the $B_{1u} + iB_{3u}$ type, is stabilized by the proximity to the ferromagnetic instability, even in the absence of long-range magnetic order. This has far-reaching consequences for the observation of residual specific heat below $T_c$~\cite{Ran19} and supports chiral currents observed recently by STM~\cite{Madhavan19}.  Further, the first principles analysis of the ``small" Fermi surface, when the $f$-$c$ hybridization is quenched by the proximity to magnetism or by applied magnetic field, reveals two quasi-two-dimensional Fermi surface sheets, in excellent agreement with the recent DFT+DMFT study~\cite{Xu19}. Knowledge of the Fermi surface geometry allows us to elucidate the nodal structure of the superconducting order parameter, which generically has nodal points along the $a$-axis, in agreement with the thermal conductivity measurements~\cite{Metz19}. In the nonunitary phase such as predicted here, these nodal points are topologically stable and the associated Majorana surface states are expected to contribute to the anomalous Hall effect. We further predict that the chiral nature of the nonunitary state should have an observable magneto-optical Kerr effect (trainable by the application of a small magnetic field), similar to that in the B-phase of UPt$_3$~\cite{Schemm14}.

\textbf{Acknowledgements.}
The author is grateful to Vaideesh Loganathan for his help with initializing the first principles calculations, and to Johnpierre Paglione, Vidya Madhavan, Youichi Yanase and Jan Tomczak for the stimulating discussions.  This research has been supported by the Robert A. Welch Foundation grant C-1818. The author is grateful for the hospitality of the Kavli Institute for Theoretical Physics, supported by the National Science Foundation under Grant No. NSF PHY-1748958, where a portion of this research has been performed.

\bibliography{UTe2}

\begin{thebibliography}{32}
\bibitem[{end({\natexlab{a}})}]{endnote1}
\bibinfo{note}{{Similar analysis has been performed for paramagnetic
  superconductor LaNiGa$_2$ in Ref.~\onlinecite{Hillier12}.}}
\bibitem[{end({\natexlab{b}})}]{endnote2}
\bibinfo{note}{{Making the assumption of equal-spin pairing, which is to say
  suppressing the $z$-component of the $\vec{d}$-vector, also has a clear
  physical meaning even in the PM phase, where the Zeeman splitting due to
  applied field may be small -- setting $d_z=0$ ensures that the transverse
  components of the orbital moment vanish, $m_x \propto \im(d_z d_y^\ast) = 0$
  and $m_y \propto \im(d_z d_x^\ast) = 0$, so that the component of the orbital
  moment along the field direction, $m_z$, is maximized.}}
\bibitem[{end({\natexlab{c}})}]{endnote3}
\bibinfo{note}{{The electron spin is not conserved because of the spin-orbit
  coupling, however the pseudospin, defined here as the projection of total
  angular momentum, is a good quantum number}}.
\end{thebibliography}
\clearpage

\newpage %\clearpage
%\section{Supplementary Materials}

\onecolumngrid

\begin{center}
	{\large \bf
	SUPPLEMENTAL MATERIAL
	} 
	
	\vspace{8pt}
	{\large \bf Stability of a Nonunitary Triplet Pairing  on the Border of Magnetism in UTe$_2$}
	
	\vspace{8pt}
    Andriy H. Nevidomskyy \\
    Department of Physics and Astronomy, Rice University, Houston, TX 77005, USA
    \vspace{4pt}
\end{center}
%%%%%%%%%% Prefix a "S" to all equations, figures, tables and reset the counter %%%%%%%%%%
\makeatletter
\setcounter{equation}{0}
\setcounter{figure}{0}
\setcounter{table}{0}
\setcounter{page}{1}
\setcounter{section}{0}
\renewcommand{\thesubsection}{S\Roman{subsection}}
\renewcommand{\thepage}{S\arabic{page}}
\renewcommand{\thetable}{S\Roman{table}}
\renewcommand{\theequation}{S\arabic{equation}}
\renewcommand{\thefigure}{S\arabic{figure}}
\renewcommand{\bibnumfmt}[1]{[S#1]}
\renewcommand{\citenumfont}[1]{S#1}
%%%%%%%%%% Prefix a "S" to all equations, figures, tables and reset the counter %%%%%%%%%%

\section{I. Details of \textit{ab initio} calculations}

The electronic band structure calculations were performed using the density functional theory (DFT) with the linearized augmented plane-waves (LAPW) as a basis, in the full-potential WIEN2K code~[\onlinecite{Wien2k}]. The generalized gradient approximation (GGA) was used to account for the exchange and correlations~[\onlinecite{PBE}]. In order to perform ``frozen-core" calculations, the U $f$-shell has been treated as core, as described in Ref.~\onlinecite{open-core}. As described in the main text, this allows one to extract the ``small" Fermi surface without the contribution of $f$ electrons without artificially ``sinking" the $f$-orbitals below the Fermi level, which is what DFT+$U$ method~[\onlinecite{SAnisimov93}] amounts to when applied to a non-spin polarized paramagnetic state. Note that in the ``frozen-core" calculation, the position of the chemical potential must be adjusted from the default one, in order to match its position relative to the all-electron calculation. In the present case of UTe$_2$ this meant shifting the chemical potential up by 1.2~eV relative to its default position.

\begin{figure}[hb!]
\vspace{-5mm}
\includegraphics[width=0.45\textwidth]{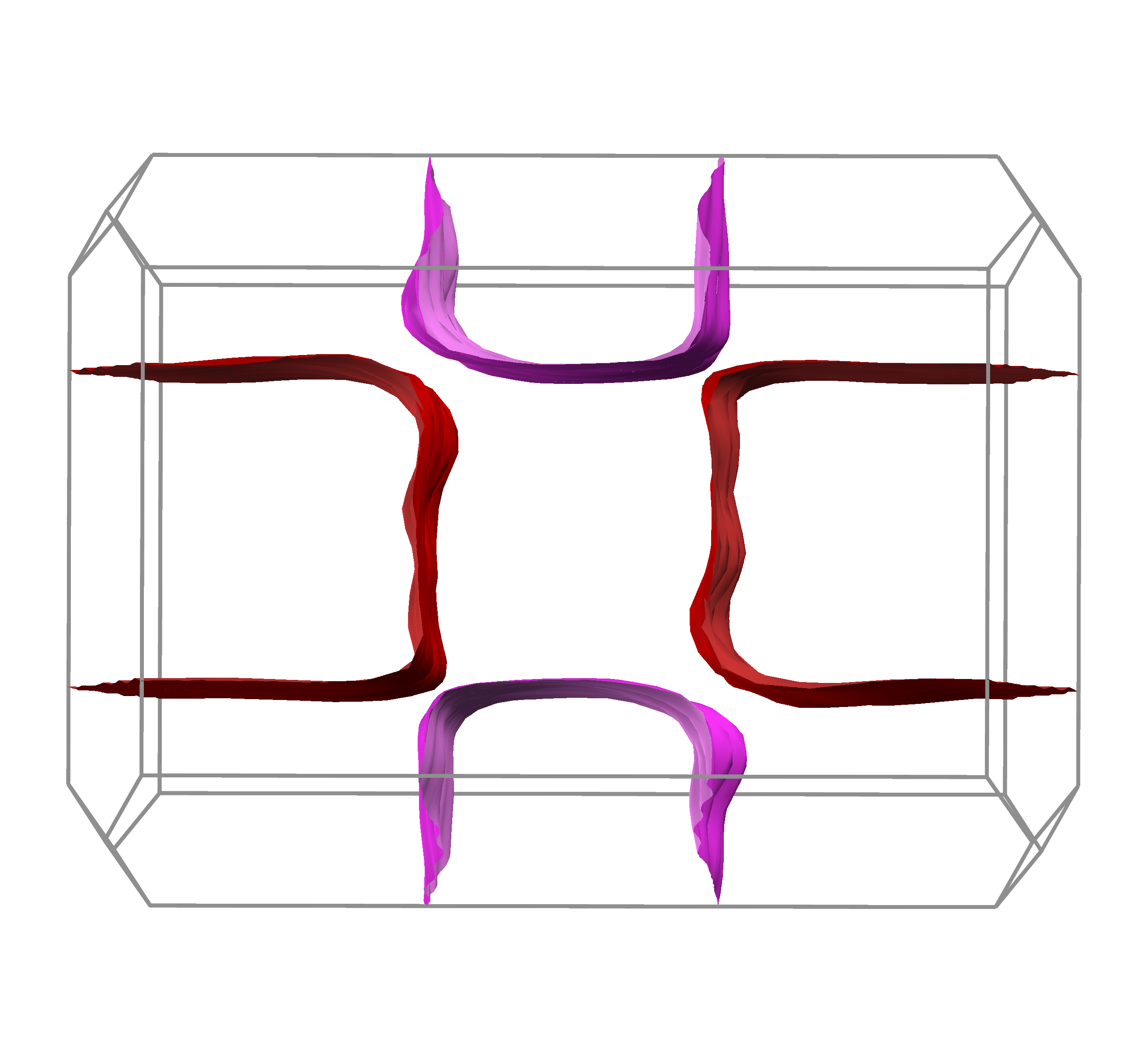}
\vspace{-4mm}
\caption{View of the Fermi surfaces of UTe$_2$ along the crystallographic $c$ direction. Very little $k_c$ dispersion is visible, and the Fermi surface is essentially two-dimensional.}
\label{fig:FS2}
\end{figure}

In order to determine the Fermi surface plots, a uniform k-point mesh of 1000 points was used, centered at $\Gamma$, and we used XCrysden~[\onlinecite{xcrysden}] for Fermi surface visualization. The resulting Fermi surfaces are quasi-two-dimensional, as seen in Fig.~\ref{fig:FS2}, and are in excellent agreement with the results of the recent DFT+DMFT calculation~[\onlinecite{SXu19}].

\vspace{-5mm}
\section{II. Pairing due to exchange of spin fluctuations}

In order to solve the gap equation, the knowledge of spin-resolved pairing potential $V_{\sigma\sigma'}$ is required. We adopt the simplified contact interaction $I$, as in the Stoner mechanism. The results for the pairing potential is obtained by summing the bubble and ladder diagrams, as explained in the seminal works on $^3$He by Berk \& Schrieffer~[\onlinecite{SBerk-Schrieffer}] and Anderson \& Brinkman~[\onlinecite{SAnderson-Brinkman}]. Here, we follow the notation of Fay \& Appel~[\onlinecite{SFay-Appel}], who give a pedagogical review and derive the results in the generic case of Zeeman-split Fermi surfaces:
\begin{align}
    V_{\sigma\sigma}(\kk,\kk') & = \left. -\frac{I^2 \chi_{\bar{\sigma}}}{1-I^2\chi_\sigma \chi_{\bar{\sigma}}}\right|_{\qq=\kk-\kk'}, \\
    V_{\up\dn}(\kk,\kk') & = -I -\left. \frac{I^2\chi_\dn\chi_\up}{1-I^3\chi_\dn\chi_\up}\right|_{\qq=\kk-\kk'} - \left.\frac{I^2\chi_{\up\dn}}{1-I\chi_{\up\dn}}\right|_{\qq=\kk+\kk'},
\end{align}
where the notation $|_\qq$ refers to the argument of the normal-state susceptibilities $\chi_\sigma(\qq,\omega=0)$, such that $\qq = \kk + \kk'$ corresponds to the ladder diagrams in the particle-particle channel, whereas $\qq = \kk - \kk'$ accounts for RPA bubbles in the particle-hole channel. Here $\chi_{\up\dn}$ is the transverse spin susceptibility (proportional to $\chi_{xx}$), whereas $\chi_\sigma$ denotes the diagonal susceptibility in a given spin channel.

It is difficult to make further progress calculating these pairing potentials without a detailed microscopic model. Fay and Appel provide key results for a jellium model with a spherical Fermi surface, and its extension to ferromagnetic superconductor UGe$_2$ has been studied by the author~[\onlinecite{SAN2005}].

In the paramagnetic phase, the calculation simplifies with $\chi_\up = \chi_\dn$, and one can further prove~[\onlinecite{SFay-Appel}] that for odd-wave pairing, the mixed-spin pairing potential $V_{\up\dn}$ precisely equals the diagonal potential $V_{\sigma\sigma}$.

\section{III. Field along $\boldsymbol{\hat{a}}$ direction}

When the field is applied along the easy $\hat{a}$ axis, or for internal magnetization developing in the ferromagnetic phase, it is convenient to rotate the coordinate frame such that $z$-axis is parallel to $a$. In this notation, the $\vec{d}$-vector of the $(\alpha B_{1u} + iB_{3u})$ nonunitary order becomes
\beq
\vec{d} = (\alpha \eta_b k_a + i\eta_b' k_c) \hat{x}  +  i\eta_c k_b \hat{y} + \alpha\eta_ak_b \hat{z}, \label{eq:dvec_a}
\eeq
and the corresponding pairing amplitudes are written as a matrix
\beq
\hat{\Delta} = i(\vec{d}\cdot\sigma)\sigma_2 = \left(
\begin{array}{cc}
    -d_x + id_y &  d_z \\
     d_z &  d_x + id_y
\end{array}
\right).
\eeq
In order to maximize the orbital moment of the Cooper pair $\textbf{m}_\text{orb} = i\vec{d}\times \vec{d^*}$ along the field direction (here chosen to lie along $\hat{z}$), we choose $d_z = 0$, as this ensures that the transverse components of $\textbf{m}_\text{orb}$ vanish: $m_x = 0 = m_y$. This is nothing but the equal-spin pairing, which transforms $\hat{\Delta}$ into a diagonal matrix. In the present case, the requirement $d_z = 0$ is satisfied by choosing $\eta_a = 0$. The resulting expression for the magnitude of the SC gap in the two spin channels are (all $\eta_i$ are taken to be real):
\beq
|\Delta_{\pm}| = \sqrt{(\alpha \eta_b k_a \mp \eta_c k_b)^2 + \eta_b'^2 k_c^2}.
\eeq
As a result, the zeros of the gap lie in the $k_c = 0$ or $k_c = \pi/c$ plane (recall that Eq.~(\ref{eq:dvec_a}) is the lowest-order expansion of a lattice-periodic function in powers of k, so that the condition $k_c = 0$ ought to be read as $\sin(k_c c) = 0$), provided that additionally 
\begin{align}
    k_b & = +\alpha (\eta_b/\eta_c) k_a, \quad \text{for zeros of } \Delta_{+} \\  
    k_b & = -\alpha (\eta_b/\eta_c) k_a, \quad \text{for zeros of } \Delta_{-} 
\end{align}
These equations, together with $k_c=0$ (or $k_c=\pi/c$) define two lines in the Brillouin zone, with the tangent of the angle with the $k_a$ axis given by $s = \pm\alpha \eta_b/\eta_c$. Looking at the top view of the Fermi surface in Fig.~\ref{fig:FS2}, and using dimensionless units for $k$ as fractions of the reciprocal vector, it is apparent that for a generic value of $|s| \lesssim 1$, the line is bound to cross the $X$-sheet of the Fermi surface (red in Fig.~\ref{fig:FS2}) at a pair of nodes, whereas a value $|s| \gtrsim 1$ would result in a pair of nodal points on the $Y$ sheet (purple). For the sake of completeness, it should be said that there is a small window of parameters where $s\approx 1$ and the line bisects the Brillouin zone without crossing either Fermi surface sheet, however this would be a very fine-tuned situation. 

We conclude that generically, applying the field (or internal magnetization) along the easy $\hat{a}$ axis would result in a nonunitary superconducting state with point nodes on either the $X$ or the $Y$ sheet of the Fermi surface, depending on the microscopic parameters of the model. This indeed agrees with the thermal conductivity data in UTe$_2$ for field $H||\hat{a}$, exhibiting behavior consistent with point nodes in the gap~[\onlinecite{SMetz19}].

\section{IV. Field in arbitrary direction}

The analysis performed above for field along the $\hat{a}$ direction and in the main text for field along $\hat{b}$ can be generalized for arbitrary field direction as follows. What one requires is a coordinate transformation such that $\hat{z}$ points along the (arbitrary) direction of the applied field. This is readily achieved by an SO(3) matrix $\hat{R}(\theta,\phi)$ parametrized by two angles. The resulting $\vec{d}$ vector in the new coordinate frame would be obtained by applying the transformation $\tilde{d} = \hat{R} \vec{d}$. To be concrete, let us consider the magnetic field in the $(bc)$ plane as in the recent experiments~[\onlinecite{SRan19-fields}]. The matrix $\hat{R}$ is then parametrized by a single polar angle $\theta$, rotating about the $\hat{a} || x$ axis:
\beq
R(\theta) = \left(
\begin{array}{ccc}
    1 & 0  & 0 \\
    0 & \cos\theta  & \sin\theta \\
    0 & -\sin\theta & \cos\theta
\end{array}
\right).
\eeq
Starting from the $\vec{d}$-vector in Eq.~(\ref{eq:dvec}) in the main text, $\vec{d} = (\alpha \eta_a k_b, \alpha \eta_bk_a + i\eta_b'k_c, i\eta_c k_b)^T$, we obtain the new vector in the rotated frame
\beq
\tilde{d} = \alpha \eta_a k_b \hat{x} + d_y \hat{y} + d_z \hat{z},
\eeq
with the maximum value of the orbital moment along the direction of the field $i(\tilde{d} \times \tilde{d})_z = 2\im(d_x* d_y^\ast)$ attained by setting $d_z = 0$:
\beq
0 = d_z = -\sin\theta\cdot\, \alpha \eta_b k_a + i[\cos\theta\, \eta_c k_b - \sin\theta\, \eta_b' k_c]
\label{eq:dz0}
\eeq
with the gap in the majority/minority spin channel given by $\Delta_{\pm} = d_x \pm d_y$, so that we obtain
\begin{align}
\re \Delta_{\pm} &= (\alpha \eta_a \pm \sin\theta\cdot\eta_c)k_b \pm \alpha\cos\theta\cdot \eta_bk_a \\
\im \Delta_{\pm} &= \pm \cos\theta\cdot \eta_b' k_c
\end{align}
From these expressions, the nodal structure of the order parameters can be analyzed for arbitrary angle $\theta$ chosen by the direction of the applied magnetic field. What this analysis implicitly assumes is that the applied field is large enough to polarize the orbital moment of the Cooper pair along the field. Whether this is the case, for a given set of microscopic parameters, is determined by minimizing the free energy in Eq.~(\ref{eq:free}) in the main text. We consider two limiting cases: \\

\noindent\textbf{1. Limit of weak magnetic field.}\\
Certainly for very weak fields, the direction of the $\vec{d}$ vector is determined not by the direction of the field, but rather that of the incipient magnetization $\mathbf{M}$ on the border of the ferromagnetic transition, as seen from the free energy expression Eq.~(\ref{eq:forb}):
\beq
\Delta F_\text{orb} = w \iint\limits_{FS}\ud^2 k\; \textbf{m}_\text{orb}(\kk)\cdot \textbf{M}.
\eeq
In UTe$_2$, the direction of the Landau magnetization $\textbf{M}$  is going to be set by the \textit{normal state magnetic anisotropy}, namely \textbf{along the easy $\hat{a}$ axis} -- this is the situations analyzed in the previous section.\\

\noindent\textbf{2. Limit of strong magnetic field.}\\
If, on the other hand, the field in the $(bc)$ plane is strong, it is expected to polarize the orbital moment of the Cooper pair $\mathbf{m}_\text{orb}$ along the field direction. It is apparent that Eq.~(\ref{eq:dz0}) cannot be satisfied to yield $d_z(\kk)=0$ for all $\kk$ for an arbitrary field angle $\theta$. Instead, consider the explicit expression for the orbital moment along the field direction ($\hat{z}$):
\beq
\langle m_\text{orb}^z \rangle 
= -2Re(\alpha)
\left(\eta_{ac}\sin\theta \iint
\limits_{FS}  k_b^2 + \eta_{ab}'\cos\theta \iint
\limits_{FS}  k_b k_c \right). \nonumber
\eeq
In the case of quasi-two dimensional Fermi surface, as follows from the \textit{ab initio} calculations (see Fig.~\ref{fig:FS2}), the last term vanishes because of integrating an odd function of $k_c$, and the free energy contribution 
\beq
\Delta F_\text{orb}(\theta) = -2w \eta_{ac} \re(\alpha)\cdot  M\sin\theta
\eeq
only depends on the component of the magnetization $M_b = M\sin\theta$ along the $\hat{b}$ direction.
Obviously, the free energy is minimized by choosing $\theta=\pi/2$, i.e. for \textbf{magnetization along the $\hat{b}$ direction}, which is why we have focused on this case in the main text of the article (note that the internal magnetization induced by the magnetic field is given by $M_b=\chi_b B$, where $\chi_b$ is the normal state susceptibility along the $b$-axis).

%\section{}

\end{document}